\documentclass{aa}
\usepackage{epsfig}
\usepackage{graphicx}
\usepackage{txfonts}
\begin{document}
   \title{The long Galactic bar as seen by UKIDSS Galactic Plane Survey}

   \author{A. Cabrera-Lavers\inst{1,2}, C. Gonz\'alez-Fern\'andez\inst{1},
           F. Garz\'on\inst{1,3}, P.L. Hammersley\inst{1}
	  \and
          M. L\'opez-Corredoira\inst{1}
          }

   \offprints{antonio.cabrera@gtc.iac.es}

 \institute{Instituto de Astrof\'{\i}sica de Canarias, E-38205 La Laguna, Tenerife, Spain\\
            \and
	    GTC Project Office, E-38205 La Laguna, Tenerife, Spain\\
                       \and
	 Departamento de Astrof\'{\i}sica, Universidad de La Laguna, E-38205 La
	 Laguna, Tenerife, Spain\\
                  }

   \date{Received XX; accepted XX}

 
  \abstract
   {Over the last decade there have been a series of results supporting the hypothesis of  the existence of a long thin bar in the Milky Way with a
   half-length of 4.5 kpc and a position angle of around 45$^\circ$. This is
   apparently a  very
   different structure from  
   the triaxial bulge of the Galaxy, which is thicker and shorter and dominates the star counts at $|l|<10^\circ$.}
   {In this paper, we analyse the stellar distribution in the inner Galaxy to see if there is clear evidence for two
   triaxial or bar-like structures in the Milky Way.}
   {By using the red-clump population as a tracer of Galactic structure, we determine the apparent morphology of the  inner Galaxy. Deeper
   and higher spatial resolution NIR photometry from the  UKIDSS Galactic Plane Survey allows us 
   to use in-plane data even at the innermost Galactic longitudes, a region where the source confusion is a
   dominant effect that makes it impossible to use other NIR databases such as 2MASS or TCS-CAIN.}
   {We show that results previously obtained with using  the red-clump giants  are confirmed with
   the in-plane data from UKIDSS GPS. There are two different structures
   coexisting in the
   inner Galactic plane: one with a position angle of 23$\fdg$60$\pm$2$\fdg$19 that can be
   traced from the Galactic Centre up to $\sim$10$^\circ$(the Galactic bulge), and other with a larger position
   angle of 42$\fdg$44$\pm$2$\fdg$14, that ends around l=28$^\circ$  (the long Galactic bar).}
   {}


\authorrunning{A. Cabrera-Lavers et al.}
   \maketitle

\section{Introduction}

The nature of the morphology of the inner Galaxy is being continously
unveiled as new and precise data on its stellar content is being
accumulated over the years. There are now little doubts, if any, of
there being a stellar bar in the Milky Way. From the pioneering work in this topic of de
Vaucouleurs (1964), using gas velocity data, to date many insights in
this direction have been produced using different methods; from the analysis of infrared (IR) surface brightness maps  
(e.g.\ Blitz \& Spergel 1991; Dwek et al. 1995) to studies of the asymmetry in the number counts  (Weinberg 1992;
Hammersley et al.\ 1994; Stanek et al.\ 1994; L\'opez-Corredoira et al. 2001),  which  both show
far  more stars at positive galactic longitudes than negative longitudes for  $l<30^\circ$, close to
the Galactic plane.

However, the exact morphology  of the inner Galaxy is still subject of
some controversy, which is also being resolved as the precise stellar
distribution is being delineated by a succesion of progessively larger
scale, depper sensitivity and higher spatial resolution NIR surveys
(TMGS, Garz\'on et al. 1993; DENIS, Epchtein et al. 1997; 2MASS, Skrutskie et al. 2006;
TCS-CAIN, Cabrera-Lavers et al. 2006; GLIMPSE,  Benjamin et al. 2003; UKIDSS, Lawrence et al. 2007).

While some authors have refered to the bar as a thick structure, around 2.5 kpc in length with a position angle of 15--30 degrees with 
respect to the Sun--Galactic Centre direction (Dwek et al. 1995; Nikolaev \& Weinberg 1997; Stanek et al. 1997; Binney et al. 1997;
Freudenreich 1998; L\'opez-Corredoira et al. 1999; Bissantz \& Gerhard 2002; Babusiaux \& Gilmore 2005), other researchers suggest that there is a 
long bar with a half length of 4 kpc and a position angle of around 45 degrees (Weinberg 1992; Hammersley et al. 1994, 2000; Sevenster et al. 1999; 
Van Loon et al. 2003; Picaud et al. 2003; Benjamin et al.\ 2005; L\'opez-Corredoira et al. 2007).

It should be noted, however, that the references supporting the shorter
bar all examine the region  at $|l|<12^\circ$ and normally off the plane, partly
due to the interstellar extinction, whereas those mentioning the long bar with the larger angle are trying to
explain counts for $10^\circ<|l|<30^\circ$. This suggests that there are probably two
different Galactic components coexisting in the inner Galaxy, instead of a single component alone with
such a complex geometry.

In this picture, the structure present in the inner Galaxy ($|l|<12^\circ$) might be interpreted as a triaxial bulge with axial ratios of 1:0.5:0.4 (L\'opez-Corredoira
et al.\ 2000, 2005) rather than a bar, and certainly not a long straight bar. Furthermore L\'opez-Corredoira et al.
(2001) showed that a triaxial bulge + disc model cannot reproduce the observed counts in the Galactic plane
 for $10^\circ<l<30^\circ$, something  also suggested by NIR photometry of red-clump stars (Nishiyama et al.\ 2005).

Along this line, Cabrera-Lavers et al. (2007a) presented a detailed
analysis of the distribution of red-clump stars in the inner Galaxy by
using deep near infrared (NIR) photometry from the TCS-CAIN survey
(Cabrera-Lavers et al. 2006). They showed that there are evidences for a
bar + bulge scenario in the innermost Galaxy. While the Galactic bulge
dominates the counts at higher latitudes, the long bar can be traced
from the in-plane data at least from l=18$^\circ$ up to l=27$^\circ$.
However, up to date, no evidences of the presence of the bulge component
were found in the Galactic plane due to incompleteness effects in the
available NIR data at those coordinates.

The intention of this paper is to repeat and complete the analysis in
Cabrera-Lavers et al. (2007a) to confirm or refute the above stated
conclusions, using the best available dataset on stellar distribution of
the central Galaxy in the NIR, the \emph{United Kingdom Infrared Deep Sky
Survey} (UKIDSS, Lawrence et al. 2007). The use of the red-clump source distribution in
the inner Galactic plane to trace the 3D morphology of the stellar
population has been performed by several authors (Hammersley et al 2000;
Stanek et al. 1994; L\'opez-Corredoira et al. 2002, 2004; Babusiaux \&
Gilmore 2005; Nishiyama et al.\ 2005, 2006b). In this paper, we will pay
close attention to the potential differences on the structural
parameters derived fron the two datasets, TCS-CAIN (Cabrera-Lavers et
al. 2007a) and UKIDSS (this work), which could then be assigned to
different stellar distribution being traced by both surveys, due to
their differences in sensitivity and spatial resolution. As it will be
shown in secs. 4.2 and 5, no significant differences are found, pointing
to the fact that the majority, or at least the most prominent section,
of the stellar content of the inner Milky Way have been already
measured. Perhaps not surprisingly, a similar results on the main
structural parameter of the stellar galactic bar can also be found in
Hammersley et al. (1994), where they analised data from the TMGS
(Garz\'on et al. 1993), with limiting magnitude of K=10.5, extended with
data from DIRBE (Boggess et al. 1992).

\section{The data: UKIDSS GPS}

The photometric data used here is taken from the UKIDSS \emph{Galactic Plane Survey} (GPS) Data Release 3. 1 x 1 degree fields 
($\Delta l$=$\Delta b$=1$^\circ$) centered on the nominal galactic
coordinates were retrieved from the UKIDSS database following the optimal query
recipe outlined in Lucas et al. (2007), up to complete the range -0.5$^\circ$$\le$l$\le$30$^\circ$, $|b|\le0.5^\circ$. By doing so we minimize the presence of
false detections (due to noise) and duplicate detections in the data. Although the authors
do not discriminate galaxies from stars in the SQL query, as we are examining areas
on the plane near the Galactic Center, the number of these sources per square degree should
be extremely low, hence any further filtering makes little difference. 

Nominal limiting magnitudes (defined as 90\% completeness limits) 
in the Vega system of UKIDSS GPS are J=19.77, H=19.00 and K=18.05, with
uncertainties of about 0.2 magnitudes (Lucas et al. 2007). 
However, for the inner in-plane fields used here confusion limited magnitudes are
slightly brighter, of the order of J=18, J=17, and K=16 at $l<20^\circ$ (and
even brighter towards  the Galactic Centre). 
These values are more than 3 magnitudes fainter
than the ones reached by 2MASS (Skrutskie et al. 2006), and 1.5 magnitudes fainter than the ones of
TCS-CAIN survey (Cabrera-Lavers et al. 2006), hence UKIDSS GPS is the most powerful tool available to date to study
these inner Galactic structures.

\section{Deriving distances from the red-clump population}
The red-clump population of giants has been shown to be an excellent distance indicator as it has 
a well defined absolute magnitude with a little dependence on age or
metallicity (Alves 2000;
Grocholsky \& Sarajedini 2002; Salaris \& Girardi 2002;
Pietrzy\'nski et al. 2003) and they are by far the most
prominent population of giants (Cohen et al. 2000; Hammersley
et al. 2000), which makes them easily identifiable in a colour-magnitude diagram (CMD). 
There are several works where this population 
has been used to determine
both densities and distances along different lines of sight with very interesting results. As examples, L\'opez-Corredoira et al.
(2002, 2004) determined some parameters for the outer and inner thin disc by using 
this population, whilst in Cabrera-Lavers et al. (2005, 2007b) the thick disc was analyzed in the same manner.

\begin{figure}[!h]
\centering
\includegraphics[width=8cm]{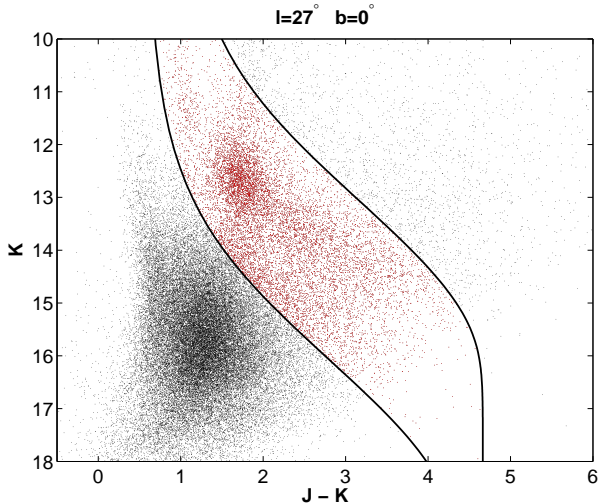}
\caption{Example of the use of the theoretical traces for the giant population by means of the SKY model. Stars isolated between both traces
(corresponding to the K0III and M0III populations) are assumed to be red-clump stars. In this particular case, note the 'bump' clearly
distinguishable at $(K,J-K)$=(12.8,1.8) that reveals the presence of an uderlying narrow structure along the line of sight.}
\label{DCM}
\end{figure}

Cabrera-Lavers et al. (2007a, hereafter CL07) also used this \emph{red-clump method} to analyze the geometry of the inner
Galaxy. Briefly, in this method the red-clump stars are first isolated in the CMDs using the SKY extinction model (Wainscoat et al. 1992) to
derive the theoretical traces for the giant population. These traces allow the red-clump stars to be extracted from the CMD, as they are assumed 
to lie between them, hence removing the contribution of dwarf stars to the counts (Fig. \ref{DCM}). Although this procedure 
includes
nearly all the K-giants in the counts, the vast majority of these stars are
red-clump stars. According to Flynn \& Freeman (1993) nearly 50\% of field red giants are in fact red-clump stars, and this percentage
 even increase to 70\% following L\'opez-Corredoira et al (2002, Fig.2). This ensures that the peak
 of the observed counts 
 coincide with the peak of the red-clump distribution, so validating the method.

The red-clump stars are then de-reddened following the relationship:

\begin{equation}
\mu=K_s-\frac{A_{K_{\rm s}}}{A_J-A_{K_{\rm s}}}[(J-K_{\rm s})-(J-K_s)_0] - M_K
\label{ke}
\end{equation}

adopting $A_{K_{\rm s}}/E_{J-K_{\rm s}}=0.68$, from Rieke \& Lebofski (1985), and assuming 
M$_K$=-1.62$\pm$0.03 and $(J-K_s)_0$=0.7$\pm$0.05 for the red-clump stars (Alves 2000; Grocholsky \& Sarajedini 2002; Bonatto et al. 2004).

Hence, we can obtain an  histogram of distance moduli along the line of
sight. This distribution would reveal the presence of any single structure, as a 'bump' of stars that
 dominates over the
background stars distribution (generally represented by a second order
polynomial). By fitting a gaussian plus the background contribution the distance modulus to the structure  can be
derived, as well as its apparent width
(defined through the sigma of the Gaussian fit):

\begin{equation}
N(m)=a+bm+cm^2+ \frac{N_{\rm RC}}{\sigma_{\rm RC}\sqrt{2\pi}}
\exp\left[-\frac{(m-m_{\rm RC})^2}{2\sigma_{\rm RC}^2}\right]
\label{nm}
\end{equation}

The full method, originally developed by Stanek et al. (1997), as well as its
uncertainties is fully described CL07 and also in Nisiyahma et al. (2006a,b) and Babusiaux \& Gilmore (2005), so we refer
the reader to those works for a further explanation of the method.

\subsection{Completeness effects}

CL07 used  NIR data from TCS-CAIN survey (Cabrera-Lavers et al. 2006), which provided
data 1-2 magnitudes deeper than 2MASS in the region $|l|<30^\circ$, $|b|<5^\circ$.
However, completeness limiting magnitudes were too bright to use the method for
in-plane fields at $|l|<15^\circ$. The results obtained in their work came from a combination of in-plane data at
$15^\circ<|l|<30^\circ$ and from off-plane data at $|l|<5^\circ$, revealing a double morphology in the
inner Galaxy, a thicker structure in the innermost Galaxy (the bulge), and a larger thin structure up to
l$=27^\circ$, constrained to the Galactic plane (the long bar).

In this work, we will follow the same method as CL07 but only use the in-plane data from the UKIDSS GPS to 
obtain the distribution of stars in 
the inner Galaxy.  This will confirm the previous result obtained using a 
combination of on- and off-plane data, as the UKIDSS GPS is notably deeper than TCS-CAIN survey in the
regions of interest.

\begin{figure}[!h]
\centering
\includegraphics[width=8cm]{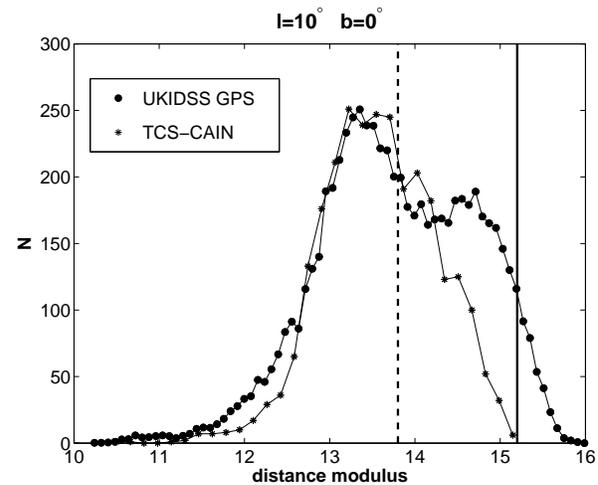}
\caption{Spatial distribution of red-clump stars obtained with either TCS-CAIN
(asterisks), or UKIDSS GPS data (points) for the
field $l=10^\circ$, $b=0^\circ$. Distributions have been normalized to the
maxima for the comparison. Vertical lines show the limiting magnitude derived in
each case. It can be seen how the completeness limit is fainter for UKIDSS data (solid
line), which is sufficient to allow  a fit to the data, whilst this was impossible
with TCS-CAIN data as the completeness limit was too close to the peak of the distribution (dashed
line).}
\label{figcompare}
\end{figure}

Figure \ref{figcompare} shows, as an example, a comparison between in-plane results at l=10$^\circ$ 
coming from  TCS-CAIN and UKIDSS GPS. It can be seen that the confusion limited magnitude is brighter in
the TCS-CAIN data, and this magnitude is too close to the peak of the distribution to perform a reliable fit to the data\footnote{In
CL07 a  minimum difference of 0.5 mag between the limiting extinction
corrected magnitude of the field and the maximum of the Gaussian fit was used to ensure 
the reliability in the
results (this is a conservative limit, what  is also used, for example, in
Babusiaux \& Gilmore 2005).}.  The
limiting magnitude is notably fainter in UKIDSS data, so in this case we can fit the distribution more
accurately. In both cases, however, the distance derived from the
distribution is nearly the same.

\section{Results: Distribution of red-clump sources in the inner Galactic Plane}
We have applied the \emph{red-clump method} to the UKIDSS Galactic Plane survey
data in steps of 1 degree for all the in-plane data at $l<30^\circ$. Results coming from the
Gaussian plus a second order polynomial fit (by means of eq. \ref{nm}) to the distance moduli distributions are
shown in Table 1, as well as the distance derived to those peaks. We have used
in each case a square of 0.25 deg$^2$ in the sky ($\Delta l$=$\Delta b$=0.5$^\circ$) centered in the nominal
galactic coordinates to extract the exact position of the peak.



\begin{table}[!h]
 \caption[]{Parameters of the Gaussian fit for 
 the distribution of the red-clump stars, and distances to the peak of these distributions for
 each in-plane ($b=0^\circ$) field used in this work.}
\label{tabla1}
      \begin{center}
    \begin{tabular}{ccccc}
\hline
$l$ ($^\circ$) & $m_{\rm RC}$ (mag) & $D$ (kpc) & $\sigma$ (mag)& $N_{\rm RC}$\\ 
\hline
    0.0     &--  &  -- &  -- & --   \\
    1.0     &--  &  -- &  -- & -- \\
    2.0 *   &13.840  $\pm$0.13 &  5.861 $\pm$ 0.350 &  0.49 $\pm$ 0.03&3114.2 \\
    3.0 *   &13.826  $\pm$0.09 &  5.823 $\pm$ 0.241 &  0.50 $\pm$ 0.08&2550.9 \\
    4.0     &14.104  $\pm$0.10 &  6.619 $\pm$ 0.304 &  0.71 $\pm$ 0.05&4169.7 \\
    5.0     &13.996  $\pm$0.12 &  6.298 $\pm$ 0.347 &  0.55 $\pm$ 0.06&2750.0 \\
    6.0     &14.142  $\pm$0.10 &  6.736 $\pm$ 0.309 &  0.95 $\pm$ 0.05&4913.1 \\
    7.0     &13.919  $\pm$0.11 &  6.078 $\pm$ 0.307 &  0.80 $\pm$ 0.06&2721.4 \\
    8.0     &13.965  $\pm$0.10 &  6.208 $\pm$ 0.285 &  0.70 $\pm$ 0.05&2149.1 \\
    9.0     &13.925  $\pm$0.08 &  6.095 $\pm$ 0.224 &  0.70 $\pm$ 0.04&2001.3 \\
   10.0     &13.633  $\pm$0.08 &  5.328 $\pm$ 0.205 &  0.85 $\pm$ 0.04&3224.5 \\
   11.0     &14.112  $\pm$0.10 &  6.643 $\pm$ 0.305 &  0.50 $\pm$ 0.05&1371.0 \\
   12.0     &14.192  $\pm$0.12 &  6.892 $\pm$ 0.380 &  0.45 $\pm$ 0.06&1557.4 \\
   13.0     &14.027  $\pm$0.10 &  6.388 $\pm$ 0.293 &  0.45 $\pm$ 0.05&969.7  \\
   14.0     &14.154  $\pm$0.10 &  6.773 $\pm$ 0.311 &  0.40 $\pm$ 0.05&530.7  \\
   15.0     &14.114  $\pm$0.11 &  6.649 $\pm$ 0.336 &  0.40 $\pm$ 0.06&766.5  \\
   16.0     &14.205  $\pm$0.08 &  6.934 $\pm$ 0.255 &  0.40 $\pm$ 0.04&676.8  \\
   17.0     &13.458  $\pm$0.08 &  4.915 $\pm$ 0.180 &  0.60 $\pm$ 0.04&1320.2 \\
   18.0     &14.039  $\pm$0.14 &  6.423 $\pm$ 0.413 &  0.53 $\pm$ 0.07&1884.1 \\
   19.0     &13.409  $\pm$0.15 &  4.806 $\pm$ 0.331 &  0.69 $\pm$ 0.08&1334.1 \\
   20.0     &13.813  $\pm$0.05 &  5.789 $\pm$ 0.133 &  0.50 $\pm$ 0.03&1766.4 \\
   21.0     &14.024  $\pm$0.12 &  6.379 $\pm$ 0.352 &  0.45 $\pm$ 0.06&1163.5 \\
   22.0     &13.956  $\pm$0.06 &  6.183 $\pm$ 0.170 &  0.65 $\pm$ 0.03&2310.2 \\
   23.0     &--  &  -- &  --   \\
   24.0     &--  &  -- &  --   \\
   25.0     &13.801  $\pm$0.13 &  5.757 $\pm$ 0.344 &  0.35 $\pm$ 0.06&972.4\\
   26.0     &13.786  $\pm$0.10 &  5.717 $\pm$ 0.263 &  0.52 $\pm$ 0.05&818.3\\
   27.0     &13.791  $\pm$0.05 &  5.730 $\pm$ 0.131 &  0.45 $\pm$ 0.02& 2016.9\\
   28.0     &13.830  $\pm$0.11 &  5.834 $\pm$ 0.295 &  0.60 $\pm$ 0.05& 1301.7\\
\hline
 \end{tabular}
\end{center}
$^{*}$ The results coming for this field are no reliable due to incompleteness.
\end{table}

\begin{figure}[!h]
\centering
\includegraphics[height=4.365cm]{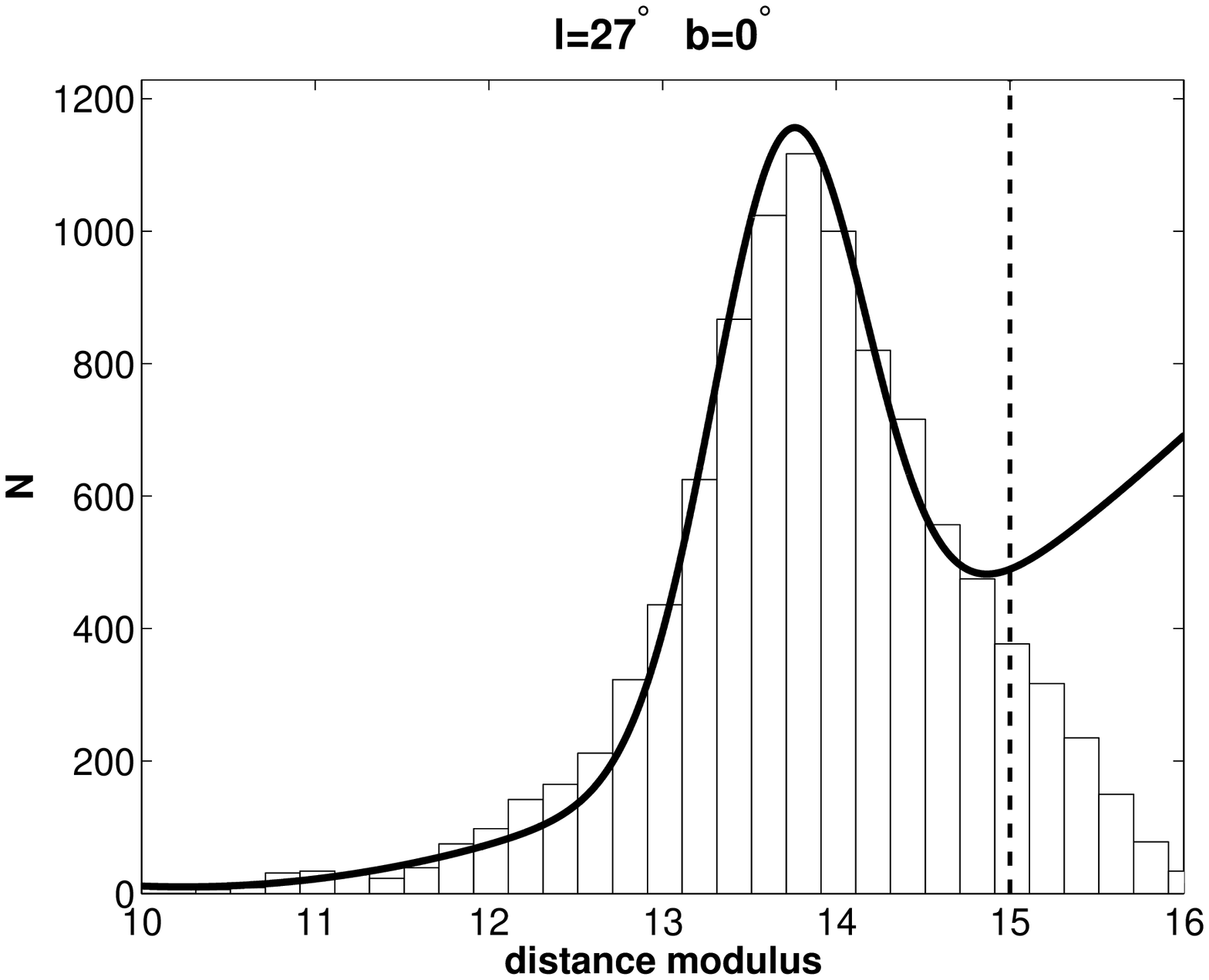}
\includegraphics[height=4.365cm]{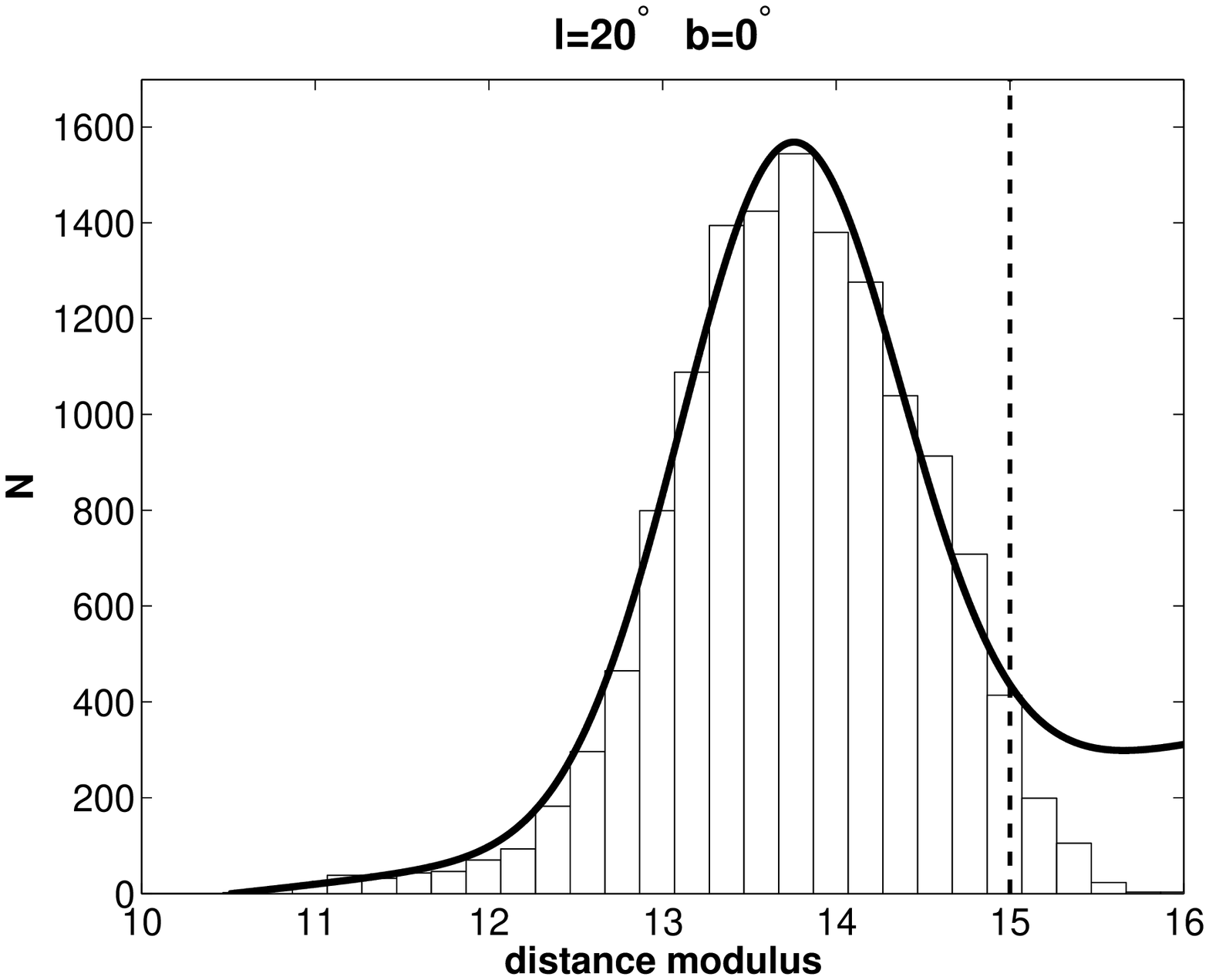}
\includegraphics[height=4.365cm]{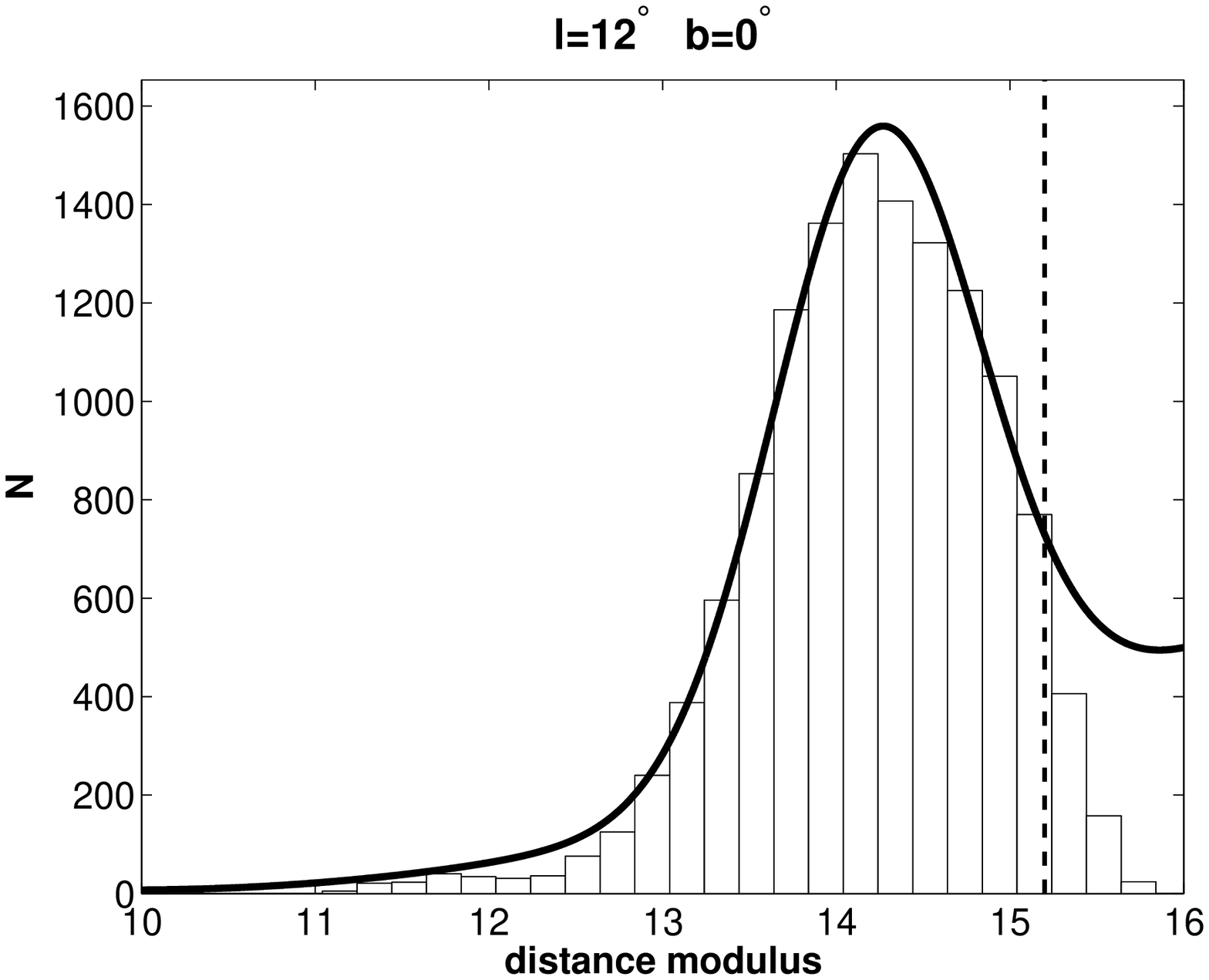}
\includegraphics[height=4.365cm]{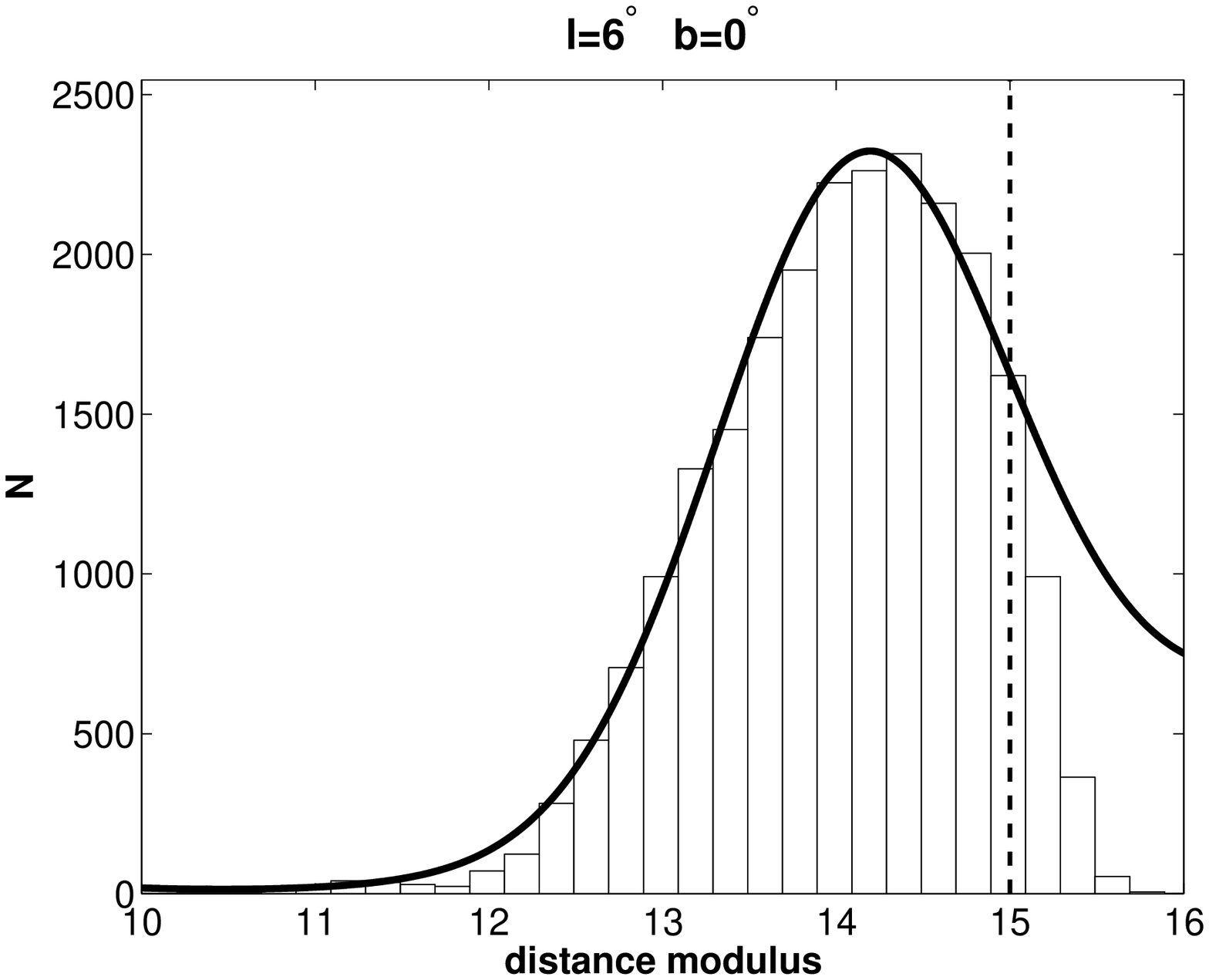}
\includegraphics[height=4.365cm]{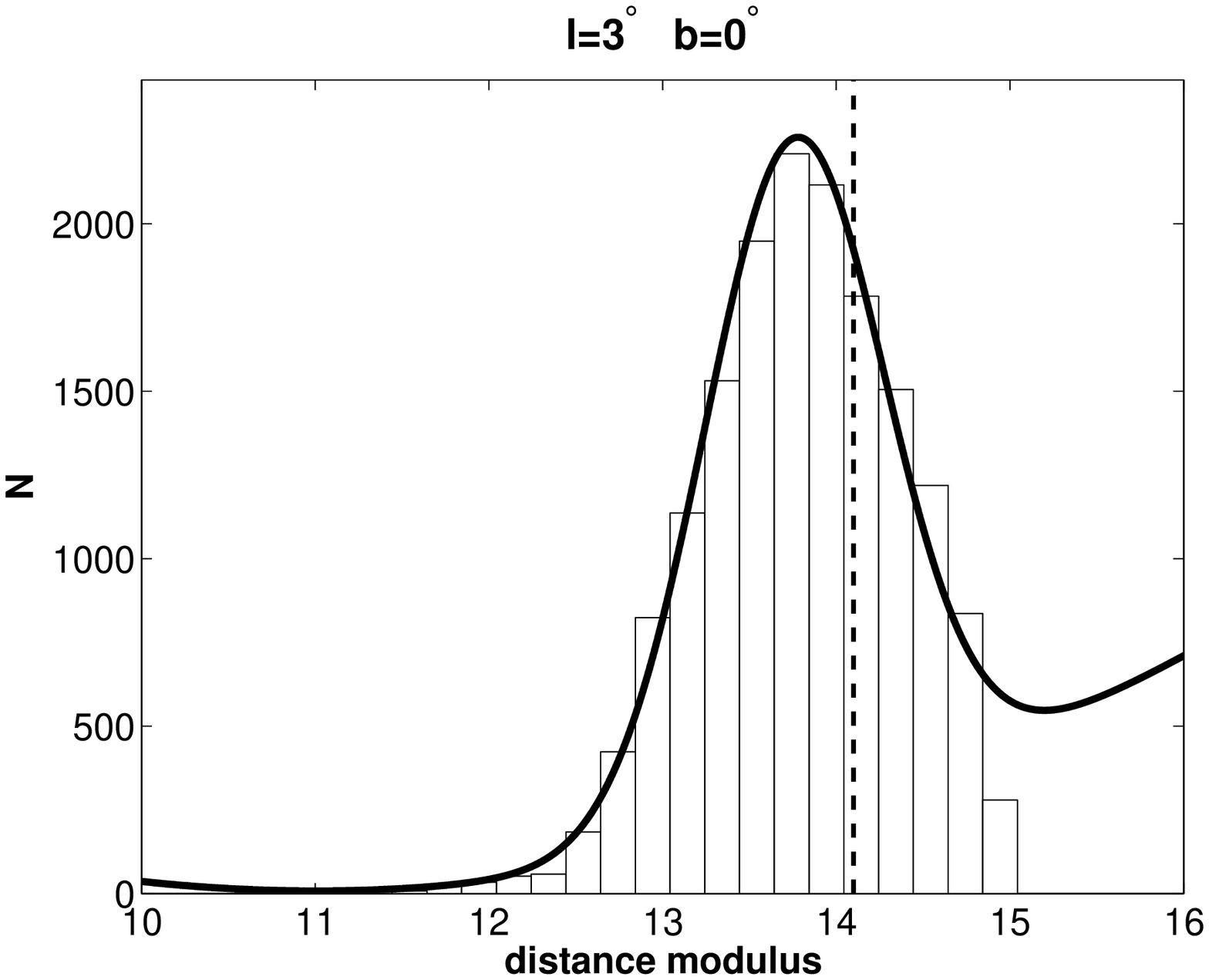}
\caption{Histograms of the distance modulus (using 0.2 mag bins) in five of the fields used in this work. 
Fits of eq.\ \ref{nm} to the histograms are shown as
solid curves, while vertical dashed lines show the limiting extinction corrected magnitudes in each field.} 
\label{fits}
\end{figure}

Figure \ref{fits} shows the fits to the de-reddened distributions of red-clump stars for five
different fields analyzed in this work. In all the cases, the 'bump' in the distribution is evident, and its location changes
depending on the longitude of  field (as the distance to the underlying structure varies). Also,
 a broadening in the distribution for the innermost fields respect to the field at l=27$^\circ$ can be noted. This
is further evidence of there being a difference between l$<10^\circ$ and l=27$^\circ$, and this will be discussed later.


There are some cases where no reliable fit was obtained due to incompleteness. 
For the innermost fields ($l\leq1^\circ$) the limiting 
extinction-corrected magnitudes are too faint to reach the 'bump' in the
red-clump distribution, while for $l=2^\circ$ and $l=3^\circ$ the peak in the red-clump
distribution is too close to the limiting extinction corrected magnitude to ensure the reliability of the fit (see for
example, the lower panel in Fig. \ref{fits}, where the field at $l=3^\circ$ is shown). Therefore
the results from  these four fields (l$\le$3$^\circ$) have been ignored for the present
analysis. 

\begin{figure}[!h]
\centering
\includegraphics[width=8cm]{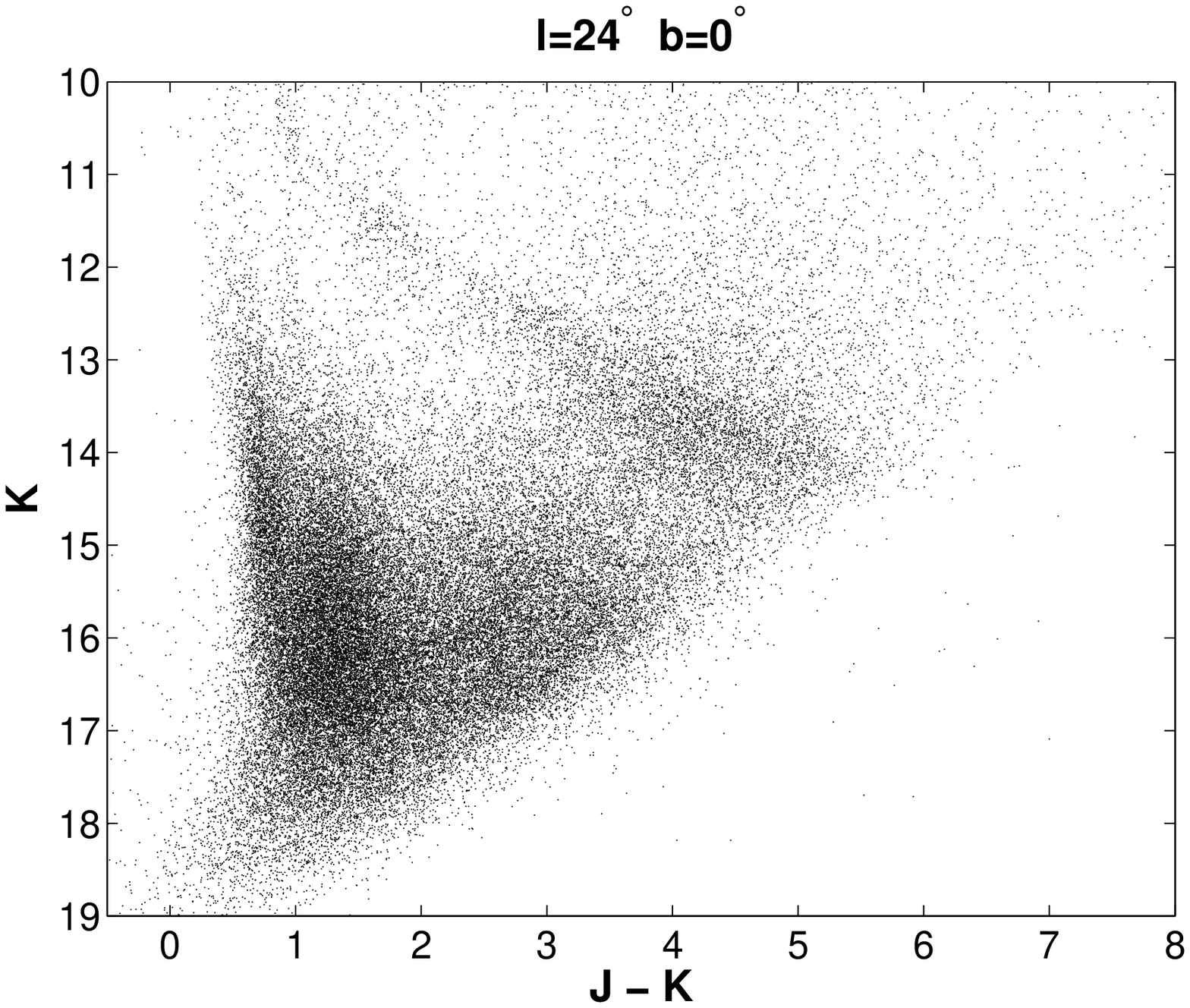}
\includegraphics[width=8cm]{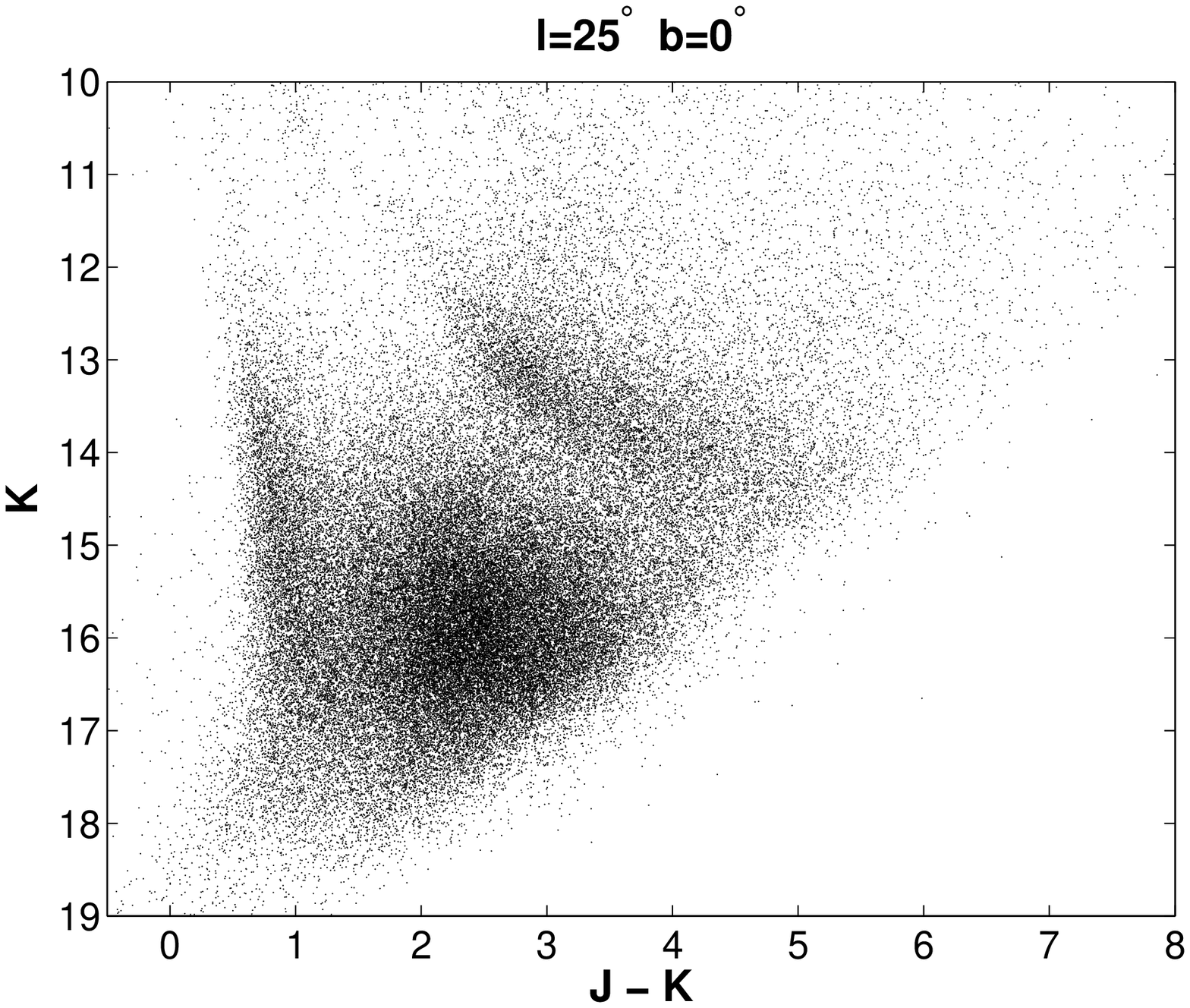}
\caption{CMDs corresponding to the in-plane fields l=24$^\circ$  (above) and l=25$^\circ$ (below). It can be noted how there is 
no apparent morphologies in
the giant branch at l=24$^\circ$, due to the high extinction present in this region.}
\label{dcm_ext}
\end{figure}

There are also two fields, at l=23$^\circ$ and l=24$^\circ$, where there is no
apparent 'bump' in the red-clump distribution, due to an effect of extinction. At these
coordinates it is well known that the extinction is very high, even in the mid-infrared
(see Fig. 1 in Benjamin et al. 2005). To illustrate this, Fig. \ref{dcm_ext} shows a comparison between the two adjacent fields at l=24$^\circ$ and l=25$^\circ$. 
The
number of stars present in both fields is clearly different even though the sky area covered is
the same. Also, the structure in the red giant branch that can be clearly noted in the l=$25^\circ$ CMD,
running from $(J-K,K)$=(2,13) to $(J-K,K)$=(4.5,14) can be also observed in the l=$24^\circ$ CMD, but appearing approximately at
$(J-K,K)$=(4,14), a value that is too close to the completeness limit of the field. Probably, a
CMD one or two magnitudes deeper is needed to observe this structure properly, at least at those coordinates.

\begin{figure}[!h]
\centering
\includegraphics[width=8cm]{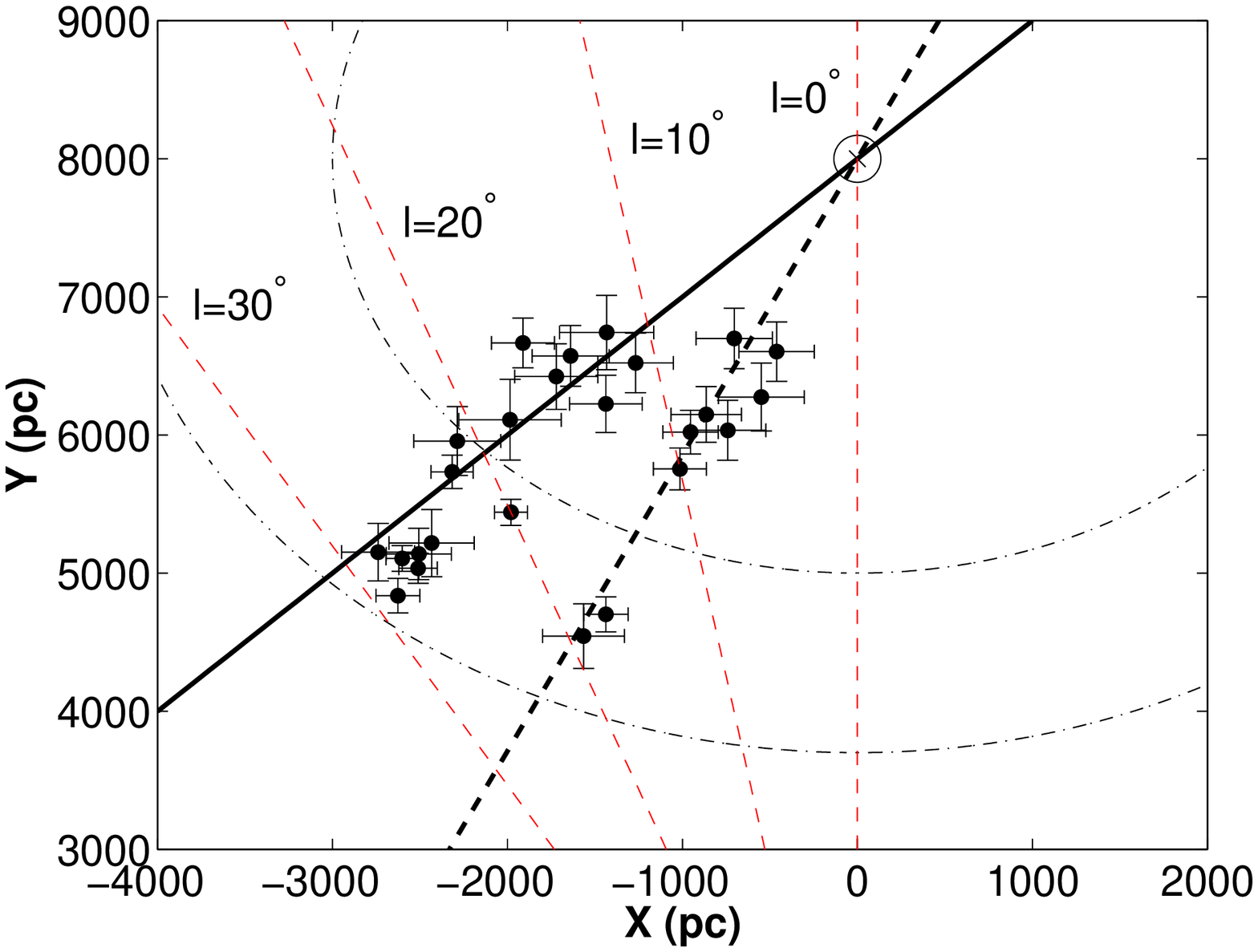}
\caption{Spatial distribution of red-clump giants maxima in the 
$XY$-plane, with the Sun at (0,0) and the Galactic Center at (0,$8000$ pc) marked with a big circle. Two possible configurations for
the observed distribution are also shown: a feature with a position angle of
45$^\circ$,  and another one with a
position angle of 25$^\circ$
(dashed line). Dot-dashed lines define two circles with radii 4.5 kpc and 3 kpc, respectively, while different lines of sight
towards the inner Galaxy are also shown in intervals of 10$^\circ$ in longitude. Error bars have been estimated from the distance
uncertainties, assuming these divided equally in both the $X$- and $Y$-axes.}
\label{b0}
\end{figure}

Fig. \ref{b0} shows the red-clump distribution in the XY plane of the Galaxy obtained from the
results presented in Table 1. There is a clear change in the trend of those data at
$\sim$l=10$^\circ$. For the innermost fields (l$<$10$^\circ$), the points lie very close to the expected geometry
for the triaxial bulge, with a position angle respect to the Sun-Galactic Centre line of 23$\fdg$60$\pm$2$\fdg$19, in agreement with the
estimates for this component (Dwek
et al. 1995; L\'opez-Corredoira et al. 2005; Stanek et al. 1997; Babusiaux \& Gilmore 2005). For the outer fields ($l>10^\circ$), the data are well distributed
along the expected geometry for the long thin bar, deriving a position angle of
42$\fdg$44$\pm$2$\fdg$14 for this feature (Hammersley 2000; Benjamin et al. 2005; CL07). For l$>$28$^\circ$, there is not
apparent underlying structure in the red-clump population, obtaining only a continuous increase in the counts as it correponds with a
typical disc field. This places the end of the long bar at approximately l=28$^\circ$, coincident with the previous result of CL07.

There are two points, those corresponding to l=17$^\circ$ and l=19$^\circ$ that lie far from the expected location for this
 longer structure that are observed
near the structure with a smaller position angle instead. The higher dispersions obtained in those fields compared to the rest of in-plane fields at 
$l>10^\circ$  (see last column in Table 1) could be a evidence that these fields are significantly  affected by extinction (although a reliable fit can be performed in
terms of completeness limiting magnitudes). A similar result was obtained in CL07 for $(l,b)$=(20$^\circ$,-0.5$^\circ$), and it is well 
known that there are a major star formation region close to those coordinates (Garay et al.\ 1998; Sridharan et al. 2005). The effect
of the stellar ring can not be even discarded, as the position obtained for the red-clump maxima in those fields is coincident with the
expected for this component (L\'opez-Corredoira et al. 2001). In any case, the over-density associated with the long Galactic bar 
is apparently somewhat inhomogeneous,  as was previously noted by Picaud et al.\ (2003) at  $l=20^\circ$ and $l=21^\circ$.

It has to be noted that the distances of the maxima obtained via red-clump fitting are not coincident with the distances of the 
density maxima along the line of sight. In
CL07 (\S 6.2) and L\'opez-Corredoira et al. (2007, Appendix A) it was demonstrated how the position angle derived with the red-clump method differs more
respect to the real one as the thickness of the underlying
structure increases. For the Galactic bar, that is a long and thin structure, differences are of the order of the own uncertainties of
the method, hence the position angle derived for this structure is nearly coincident with the real one. For the triaxial bulge, 
the difference is higher, producing a variation of 8-10$^\circ$ respect to the real position angle of the bulge. Hence, the position angle
for this latter component might be around 13-15$^\circ$, coincident with the value previously obtained in CL07.

\subsection{Thickness of the components}
\begin{figure}[!h]
\centering
\includegraphics[width=8cm]{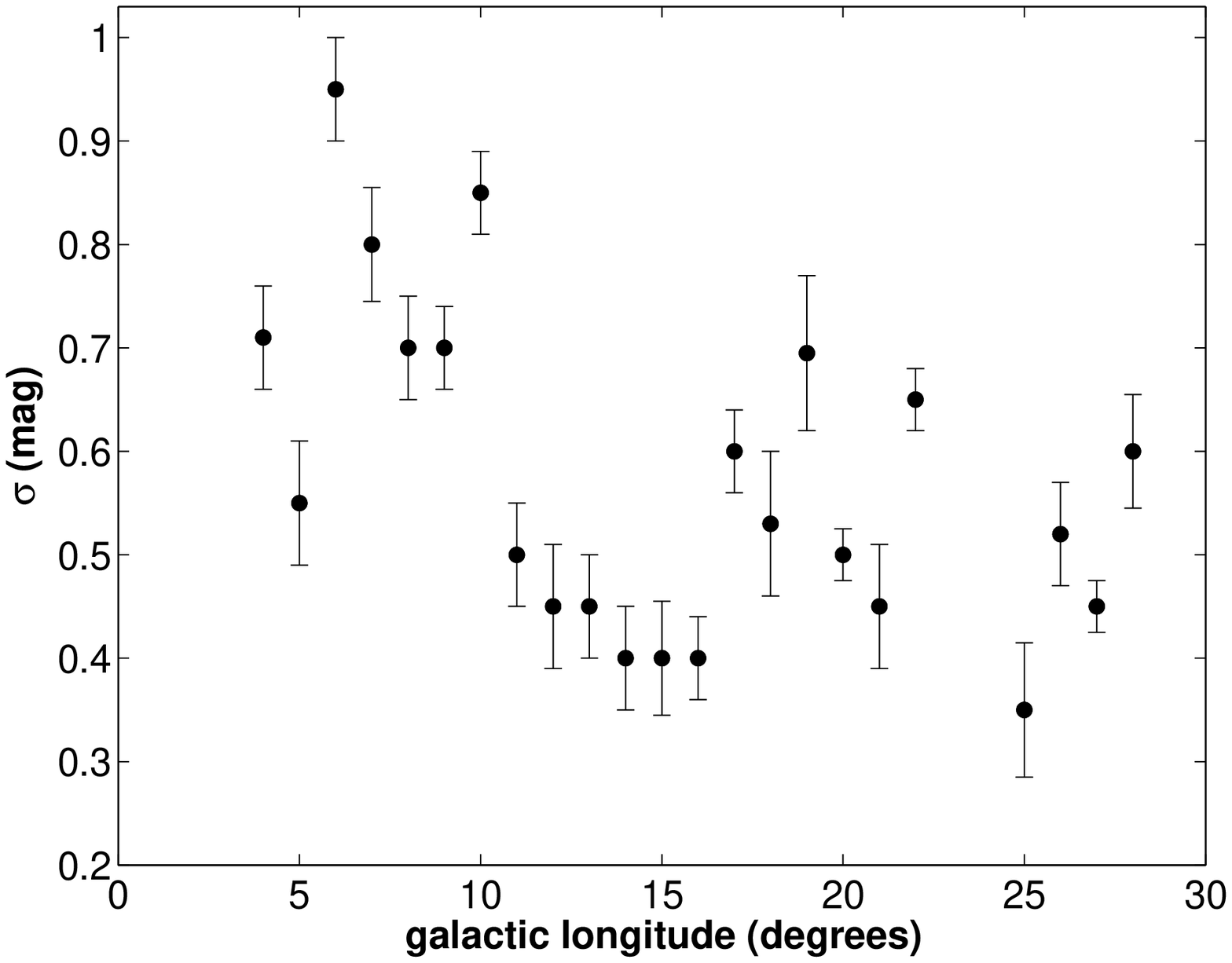}
\caption{Variation in the width of the red-clump distribution with the galactic longitude. }
\label{sig}
\end{figure}

The dispersion in distance obtained for each field (represented by $\sigma$ in Table 1) also shows a dependence with the Galactic longitude
(see Fig. \ref{sig}). It is clear that for the innermost fields
(l$<$10$^\circ$) the peak of the
red-clump distribution notably broadens with respect to the one obtained for the larger Galactic longitude fields (this effect can be also
observed in Fig. \ref{fits}). We obtain a mean
value of $<\sigma>$=0.751 mag (with a standard deviation of 0.128 mag) for l$\le$10$^\circ$, while for l$>$10$^\circ$ this average is
notably smaller, with $<\sigma>$=0.496 mag (and a standard deviation of 0.098). This is indicative that the structure we are observing
at l$<$10$^\circ$ is different to the one that dominates at l$>$10$^\circ$, as the width of the distributions are different by more than 
2.5$\sigma$ over the average. The same effect was also noted by CL07 and L\'opez-Corredoira et al. (2007, Fig. 4). If we perform a linear fit
to those points in Fig. \ref{sig} we obtain:  $\sigma$ = 0.74 - 0.01 $l$, not very far from the $\sigma$ = 0.64 - 0.004 $l$ obtained in
L\'opez-Corredoira et al. (2007). As expected, the triaxial bulge is notably thicker along the line of sight than the long Galactic bar, hence the width of the red-clump peaks
for the fields
associated to this component might be higher to those of the long bar, that is assumed to have a half-width of about 0.6 kpc
(L\'opez-Corredoira et al. 2007). CL07 also
obtained that for in-plane fields at $l>15^\circ$ there were very homogeneous range of dispersions in the $\sigma$ of the distribution with an
average of $\sigma$=0.49 mag, nearly coincident with those obtained here, which reinforces our conclusions.

There is also a notable increase in the number of red-clump sources for the innermost fields (see last column in Table 1). 
The number of red-clump stars at $l<10^\circ$ are on average nearly three times
the values obtained at $l>10^\circ$. This indicates that in the region where the contrast between the triaxial bulge and the long bar 
is higher, the long bar lies completely embedded in the bulge counts, making impossible to note its
contribution to the histogram counts (see, for example, Fig. \ref{figcompare} and middle panel in Fig. \ref{fits}).


\subsection{Comparison with CL07}

\begin{figure}[!h]
\centering
\includegraphics[width=8cm]{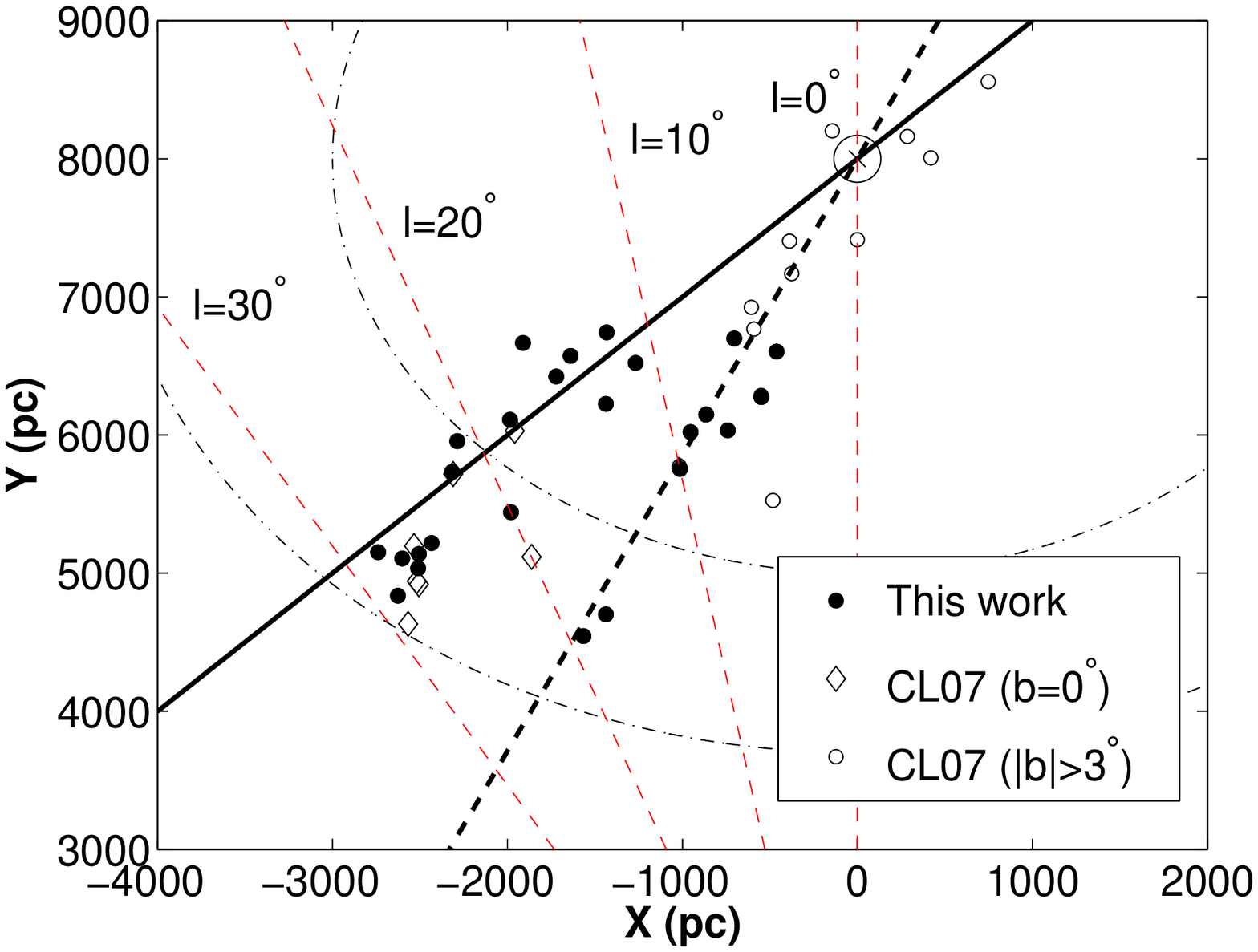}
\caption{Same as Fig. \ref{b0}, but now including the results obtained in CL07
for a combination of in-plane and off-plane data from TCS-CAIN survey.}
\label{b0_CL}
\end{figure}

We show in Fig. \ref{b0_CL} the results obtained in CL07 for the
combination of in-plane data at $l>15^\circ$ and off-plane data for $l<10^\circ$, to compare with those of Fig. \ref{b0}. The
agreement between both sets of data is evident. While off-plane results  
in CL07 are coincident with the in-plane data of UKIDSS GPS in the innermost fields, data
for l$>$10$^\circ$ in UKIDSS GPS coincides with the in-plane fields used in CL07. 

\begin{figure}[!h]
\centering
\includegraphics[width=8cm]{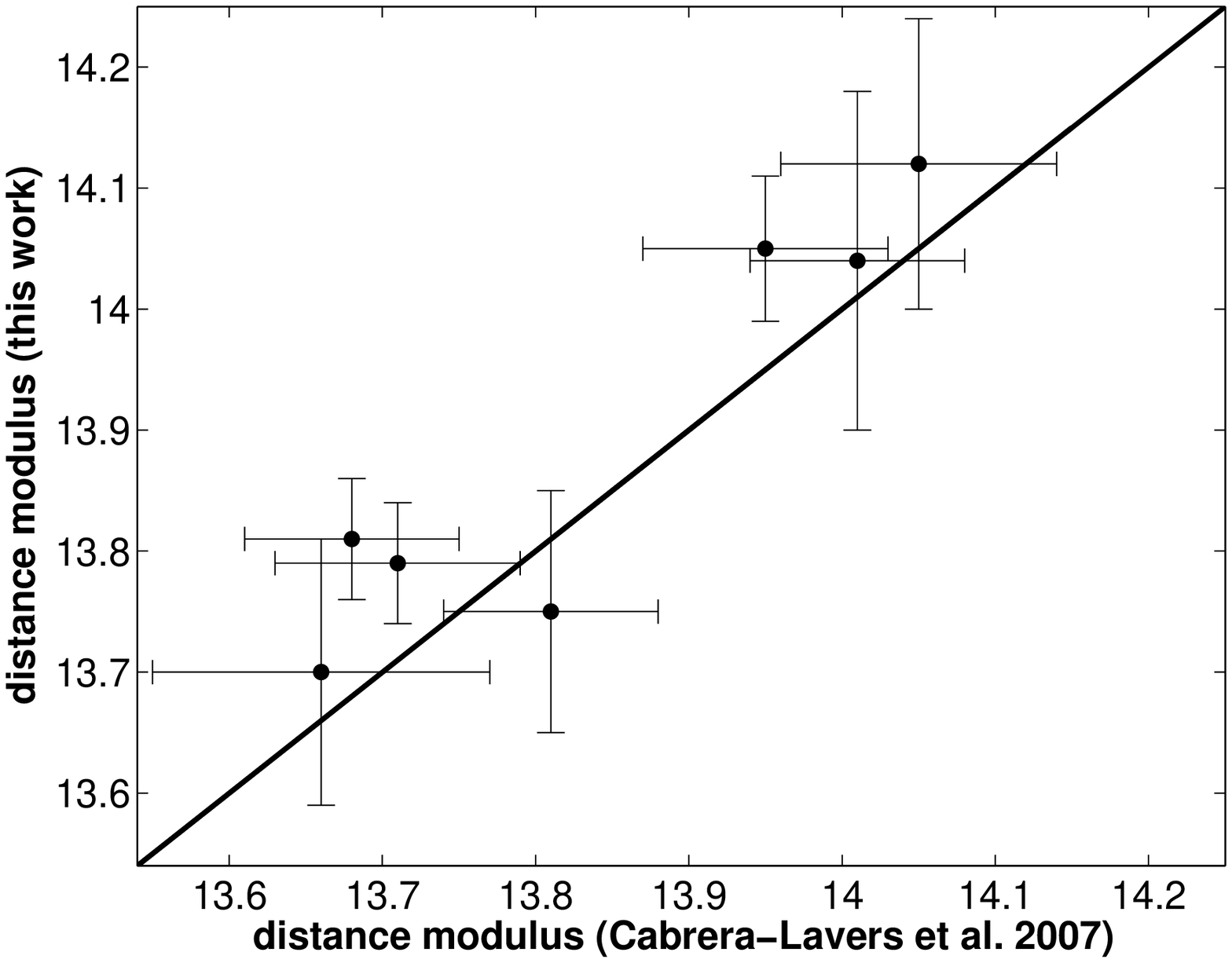}
\caption{Comparison between the distance modulus obtained in this work
data with that obtained in CL07 for the same
fields. The solid line shows the 1:1 ratio. There is a offset of 0.05 mag between both data sets related to the different characteristics
of the K filters used in both surveys.}
\label{mu}
\end{figure}

The mean difference in distance moduli between the in-plane fields in common analyzed in
this work and in CL07 is $\Delta \mu$=0.05 $\pm$ 0.02 mag (see Fig. \ref{mu}). By comparing the photometry of TCS-CAIN data and 
UKIDSS GPS in several test fields we obtained mean differences  (UKISS-TCS) of $\Delta J$=0.015 mag ($\sigma_J$=0.183) and $\Delta K$=0.167
mag ($\sigma_K$=0.256). The larger difference in the K data is probably due to the different characteristics of the K filter 
used in both surveys. In any case, a difference in distance modulus of 0.05 mag yields to a difference in distance of only 150 pc
on average, of the order of the distance uncertainties for each individual field. Hence, the results coming from CL07 and this work
can be considered equivalents in terms of photometry, and are comparable.

\subsection{Systematic errors in the absolute magnitude of the red-clump population}

\begin{figure}[!h]
\centering
\includegraphics[width=8cm]{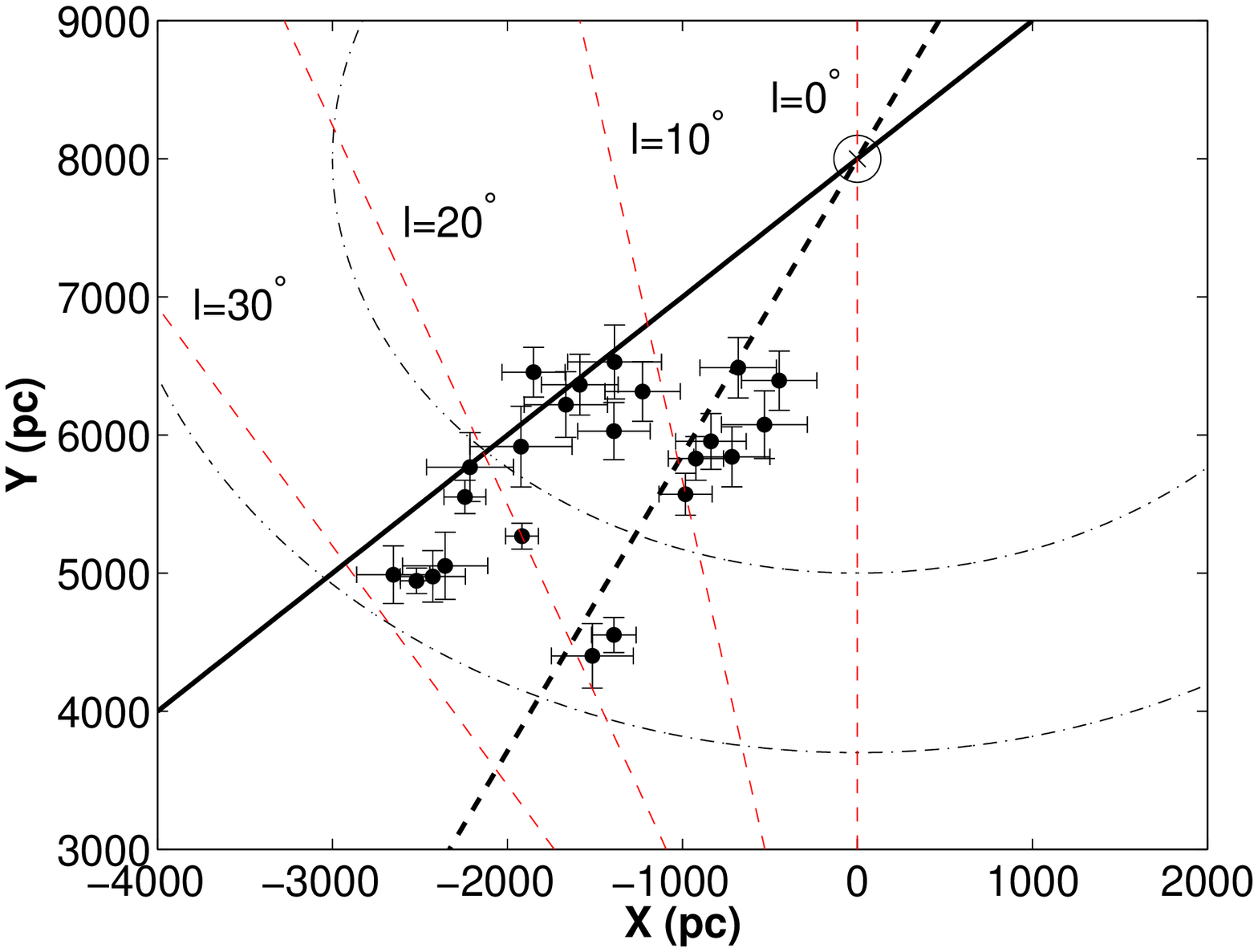}
\caption{Same as Fig. \ref{b0}, but assuming an absolute magnitude of \mbox{M$_K$=-1.54} for the red-clump stars.}
\label{b0_mk}
\end{figure}

A recent recalculation of the absolute magnitude for red-clump stars by Groenewegen (2008) gives a value 
slightly different to that is
assumed in this work ($M_K$=-1.62$\pm$0.03). It was found that the absolute magnitude determination for red-clump 
stars via \emph{Hipparcos} data was biased to brigther stars, and the real 
value was slightly fainter: $M_K$=-1.54$\pm$0.04. This adds a systematic uncertainty of 0.08 mag in the
values obtained along this work, which translates in differences up to 0.15 kpc for the closest points (D=5 kpc) and 0.20 kpc for the
farthest ones (D=6.5-7 kpc). This possible effect, however, does not affect the main conclusions of the work presented here, as the 
geometry observed in the inner Galaxy remains the same (Fig. \ref{b0_mk}). The measured points move closer nearly the same
 distance along the 
line of sight, and the two observed structures can be traced unambiguosly. Values for the position angles are also slightly different, as the
distance to the Galactic Centre has been kept on 8 kpc:
39$\fdg$97$\pm$2$\fdg$22 for the long bar and 20$\fdg$93$\pm$2$\fdg$61 for the triaxial bulge. However they are still clearly different
 and the two-component geometry is evident.

\section{Conclusion}
By using the deepest wide area NIR data available to date for analyzing the inner Galactic Plane, we find that 
there are two different structures coexisting in the inner 4 kpc of the Milky Way: From the Galactic Centre to l=10$^\circ$, 
the distribution of red-clump sources traces a thicker structure with a position angle with respect to the Sun-Galactic Centre line of 
23$\fdg$60$\pm$2$\fdg$19, while for l$\le$10$^\circ$ the red-clump sources follow a more inclined structure 
that can be traced up to l=28$^\circ$ with a position angle of 
42$\fdg$44$\pm$2$\fdg$14, that ends around l=28$^\circ$. 

This result corroborates the previous hypothesis of Hammersley et al. (2000) that the Milky Way presents a inner triaxial
 bulge together with a long thinner structure, that appears  as a long Galactic bar, with semi-length of about 4.5 kpc 
 from the Galactic Centre. There are also evidences that the inner triaxial bulge has a 
 'boxy' morphology (L\'opez-Corredoira et al. 2000,
 2005), but nothing in that sense can be said from the analysis presented here. 
 
Previous works in the
same direction (CL07 and L\'opez-Corredoira et al. 2007) could not analyze the inner Galactic Plane due to the
shallowness in their NIR data. For this reason, they had to use a combination of on and off plane data to reach the same conclusions as ours.
Some other researchers also pointed to this double morphology in the Milky Way, and suggested that the inner Galaxy is
dominated at higher latitudes by the Galactic Bulge, 
while the star counts in the plane are dominated by the long bar who dominates the counts (Sevenster et
al. 1999). The result presented here confirms that the same picture can be observed close to the Galactic Plane.

This suggested  scenario of a triaxial bulge + long bar for the  Galaxy is plausible as many galaxies have triaxial bulges contained 
within a primary stellar bar (Friedli et al. 1996), and it is well supported by recent N-body simulations about secular galactic
evolution\footnote{In fact, the ratio of sizes between the semi-major axis of the long bar and the major axis of the triaxial bulge in the
simulations is 1.5, nearly coincident what it is observed in Our Galaxy.}. (Athanassoula 2005; Athanassoula \& Beaton 2006) and also by
NIR observations in external galaxies (Beaton et al. 2005). Boxy bulges are in fact a part of the long bar (Athanassoula 2005), and this
is what is seen in the  Milky Way.



\begin{acknowledgements}
We gratefully acknowledge the anonymous referee for very interesting suggestions that have notably improved
 the quality of this paper. This work is based on data from the 3rd data release of the 
 UKIRT Infrared Deep Sky Survey, UKIDSS (www.ukidss.org), which is described in detail in Lawrence et al. (2007) and Lucas et al. (2007).

\end{acknowledgements}

\end{document}